\documentclass{article}

\usepackage{microtype}
\usepackage{graphicx}
\usepackage{subfigure}
\usepackage{booktabs} %
\usepackage{color}
\usepackage{enumitem}

\usepackage{hyperref}

\usepackage[accepted]{mlsys2025}
\makeatletter
\renewcommand{\Notice@String}{Preprint. Copyright 2026 by the author(s).}
\makeatother
\hypersetup{pdfauthor={Hamidreza Alikhani Koshkak, Nazmul Takbir, Matin Khajavi, Rajat Bhattacharjya, Sangeetha Abdu Jyothi, Nikil Dutt},
  pdfsubject={LLM serving},
  pdfkeywords={LLM serving, prefill scheduling, KV cache reuse, prefix caching, retrieval-augmented generation, TTFT, tail latency, fairness}}

\mlsystitlerunning{Fairway: Load-Aware KV Cache Reuse for LLM Prefill Scheduling}
\usepackage{amsthm}
\ifdefined\lemma\else\newtheorem{lemma}{Lemma}\fi
\ifdefined\corollary\else\newtheorem{corollary}{Corollary}\fi
\definecolor{teal}{rgb}{0.0,0.4,0.0}
\providecommand{\sys}{\textsc{Fairway}}
\begin{document}

\twocolumn[
\mlsystitle{Fairway: Load-Aware KV Cache Reuse for LLM Prefill Scheduling}

\begin{mlsysauthorlist}
\mlsysauthor{Hamidreza Alikhani}{uci}
\mlsysauthor{Nazmul Takbir}{uci}
\mlsysauthor{Matin Khajavi}{api}
\mlsysauthor{Rajat Bhattacharjya}{uci}
\mlsysauthor{Sangeetha Abdu Jyothi}{uci}
\mlsysauthor{Nikil Dutt}{uci}
\end{mlsysauthorlist}

\mlsysaffiliation{uci}{University of California, Irvine, CA, USA}
\mlsysaffiliation{api}{Apiphany}

\mlsyscorrespondingauthor{Hamidreza Alikhani}{hamidra@uci.edu}

\mlsyskeywords{LLM serving, prefill scheduling, KV cache reuse, prefix caching}
\vskip 0.3in

\begin{abstract}
  {
  LLM-based applications often employ agentic workflows and retrieval-augmented generation (RAG), which build long prompts from shared context. When prefixes are shared across requests, serving engines reuse the key-value (KV) cache to skip parts of the prefill computation. Reuse is typically graded: the uncached work of requests in a queue can differ by up to $100\times$. In this setting, the order of scheduling determines the time-to-first-token (TTFT) for the queued requests. KV reuse-first scheduling, which prioritizes the requests with the longest KV cache hit, gives the lowest median TTFT but starves expensive
  requests under load spikes; first-come-first-served (FCFS) scheduling
  bounds the tail but raises the median TTFT. We present \sys{}, a prefill scheduler built on \emph{adaptive reuse prioritization}: it favors KV cache reuse while queues are short, and
    shifts toward fair allocation as requests wait longer relative to their own
    prefill time. \sys{} employs a virtual dual lane model: a KV reuse-first \emph{Fast lane} and a \emph{Fair lane}
    atop the serving engine's waiting queue that enables speeding up of cheap reuse
    requests without starving expensive ones. This
    model automatically adapts to the workload without any per-workload tuning. \sys{} lowers the tail TTFT at no cost
    to the median in transient overload. Across workloads and load levels, \sys{} matches or beats the best
    baseline, with a p99 TTFT up to $8.6\times$ lower than prior
    prefix-aware schedulers. During load spikes, it keeps the tail of FCFS at a
    $6\times$ lower median TTFT.
    
    }

\end{abstract}

]

\printAffiliationsAndNotice{}

{
\section{Introduction}
\label{sec:intro}
Retrieval-augmented generation (RAG) and agentic workflows assemble long
prompts from context shared across requests. RAG adds retrieved documents
to each prompt, as in enterprise search and document question
answering~\cite{rag, ragsurvey}, and an agent resends its history and tool
outputs on every call~\cite{mooncake}. LLM inference runs in two phases:
\emph{prefill} processes the whole prompt in one compute-bound pass,
building its key--value (KV) state and emitting the first token, and
\emph{decode} then generates the remaining tokens one at a time. These
prompts are long, so the time to first token (TTFT) is dominated by
prefill. State-of-the-art serving engines therefore cache the KV state of
prompt prefixes and skip recomputing it~\cite{vllm,sglang}. The KV cache
is finite, so the order in which waiting requests are served sets both
how long each request waits and whether its prefix is still cached when
it runs: a request
held back in the queue may lose its hit when the requests admitted before
it evict its KV blocks. In prefill--decode disaggregated
serving~\cite{distserve,splitwise,mooncake}, this order is set on each
prefill node, where a request spends most of its TTFT queueing and in
prefill.

Reuse is \emph{graded}: a request may find all, some or none of its
prefix cached, so its prefill work depends on its uncached tokens rather
than its prompt length. In Mooncake's production agentic
traces~\cite{mooncake}, most requests are partly cached, and this work
differs by $\approx100\times$ within a queue (\S\ref{sec:mot:cost}). RAG
behaves the same way: requests share some retrieved documents but not
others, so full hits that compute the prefill only for their question wait in the same
queue as partial hits that carry nearly all the prefill work.
Prompt lengths vary as well, and the mix of hits and lengths shifts as the
popularity of documents and the load change over time.

Existing schedulers handle this variation only partially.
KV reuse-first schedulers, which first serve the request with the longest
cached prefix~\cite{sglang} or the highest ratio of cached to uncached
tokens~\cite{ragcache},
give the best median TTFT but can starve expensive requests when load
rises. First-come-first-served (FCFS) bounds the tail but ignores reuse,
so cheap hits wait behind long misses. Schedulers that add fairness to a
KV reuse-first order measure it in units that ignore request size: some
count turns, giving the oldest request one turn in every
few~\cite{klpm} or sharing turns among groups of requests with similar
cache ratios~\cite{preble}, though one turn may cost tens of tokens or
thousands; other works count seconds, raising a request's priority with its
absolute wait~\cite{prefillonly}, though the same wait time weighs differently
on requests of different lengths. Figure~\ref{fig:teaser} shows the
result in a short load spike: KV reuse-first orders keep a median of
0.8\,s, but their p99 reaches 15--42\,s; FCFS keeps the p99 at 7.8\,s,
but its median rises to 5.9\,s. Each fixed rule suits some mix of hits,
lengths and load, and fails on another (\S\ref{sec:bg:sched},
\S\ref{sec:mot}).
\begin{figure}[t]
  \centering
  \includegraphics[width=\columnwidth]{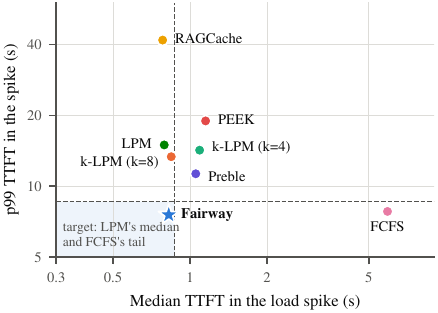}
  \vspace{-5mm} %
  \caption{Median vs.\ p99 TTFT during a 60\,s load spike on a finance RAG
  trace (SEC QA; log--log). Shaded: within 10\% of LPM's median and of FCFS's p99.}
  \label{fig:teaser}
  \vspace{-6mm} %
\end{figure}

We present \sys{}, a prefill scheduler built on \emph{adaptive reuse
prioritization}: it favors KV cache reuse while queues are short and
shifts toward fairness as queues grow and requests wait longer relative
to their own prefill time. \sys{} rests on two choices: it measures
fairness in \emph{work} characterized by a request's uncached tokens, and it measures each request's \textit{wait} relative to its own expected
prefill time. 
Two virtual lanes, which order the same
requests in two ways, form the backbone of \sys{}. The \emph{Fast lane} ranks
requests by the compute their cached prefix saves and promotes them as
their \emph{relative wait} grows, and the \emph{Fair lane} ranks them by
arrival time. \sys{} adaptively prioritizes between the lanes, balancing KV cache reuse and fairness with a bound on the work placed ahead of the oldest requests.
This balance adapts automatically with different workloads, without requiring any
mode switch or per-workload tuning.

We implement \sys{} as a drop-in hook in vLLM's scheduler~\cite{vllm}: \sys{} changes how waiting requests are selected for admission in each scheduling step. It operates orthogonally to other engine policies, such as eviction and step-size selection, and leaves those and other engine configuration unchanged. We evaluate it
on a production agentic trace~\cite{mooncake} and on RAG workloads for financial and general multi-hop QA ~\cite{mooncake,kgrag,musique,longbench,loogle,klpm,preble,peek} across a wide range of loads and prompt lengths.
We compare it with seven baselines, from FCFS and KV reuse-first policies
to schedulers that consider fairness in their admission policy~\cite{klpm,preble,peek}.

Across workloads and load levels, \sys{} matches or beats the best
baseline in every scenario. Its p99 TTFT is up to $8.6\times$ lower than that of
prior prefix-aware schedulers, and it starves no request.
 In load spikes, \sys{} keeps the tail of FCFS while
achieving a $6\times$ lower median TTFT, by letting cheap cache hits go
first while bounding how far they can push expensive misses back.
On mixed-length workloads, it cuts the p99 TTFT of short
requests by about a third relative to every baseline that does not starve
requests.

In summary, we make the following contributions:
\begin{itemize}[itemsep=0pt, topsep=0pt] %
  \item A characterization of graded prefix reuse: We show that in agentic
  traces and RAG workloads, prefill work within a queue differs by
  $\approx100\times$, depending on both the cache state and the prompt length
  (\S\ref{sec:mot:cost}).
  \item An analysis of why existing schedulers fall short: KV reuse-first
  orders starve expensive requests under load spikes, FCFS raises the
  median, and fairness counted in turns or in absolute wait time ignores
  how much work each request carries
  (\S\ref{sec:mot:cost}--\S\ref{sec:mot:bound}).
  \item \sys{}, a prefill scheduler built on \emph{adaptive reuse
  prioritization}, which measures fairness in work and each request's wait
  relative to its own prefill time. Two virtual lanes, merged by the work
  each has scheduled, favor KV cache reuse while queues are short and
  bound the work served ahead of the oldest request as queues grow,
  without per-workload tuning.
(\S\ref{sec:design}).
  \item Evaluations against seven baselines on a production agentic trace and three RAG benchmarks under a wide range of workload characteristics showing that \sys{} stays robust where the best baseline changes,
  with one setting throughout (\S\ref{sec:eval}).
\end{itemize}
}

\vspace{-9pt}
\section{Background and Related Work}

\label{sec:bg}
\subsection{Prefill Serving}
\label{sec:bg:prefill}

Serving a request consists of two phases. \emph{Prefill} processes the whole prompt
in one pass, writes its key--value (KV) cache, and emits the first token;
\emph{decode} then generates the remaining tokens one at a time. The
time-to-first-token (TTFT) is therefore the time a request waits in the
engine's queue plus the time of its prefill. Engines such as vLLM~\cite{vllm} run in \emph{steps} with continuous
batching~\cite{orca}. At every step the scheduler first continues the
requests that are already running, then admits waiting requests in queue
order while the step's token budget lasts. With chunked
prefill~\cite{sarathi}, a step carries at most $B$ prompt
tokens,\footnote{vLLM 0.16.0 uses $B=2{,}048$ for its API server on GPUs
with less than 70\,GB of memory and $8{,}192$ on H100-class GPUs.} so a
long prompt is split over several steps.

In prefill--decode (P/D) disaggregated serving~\cite{distserve,splitwise,mooncake},
prefill runs on dedicated nodes that hand the KV cache to decode nodes as
soon as the first token is computed, so a request spends most of its TTFT
queueing and in prefill on the prefill node. TTFT is the first latency a
user sees in interactive agent and RAG applications, and we target this
node.

The engine has other levers, which \sys{} leaves unchanged and composes
with: the step budget~\cite{sarathi}, eviction, preemption, fairness
across clients~\cite{vtc,dlpm}, and routing across instances by cache
locality and load~\cite{mooncake,llumnix,preble}. We study one:
\emph{which waiting request the engine admits next}.

\subsection{Prefix Caching}
\label{sec:bg:prefix}

Prefix caching lets a request reuse the KV cache that an earlier request
computed for the same leading tokens~\cite{vllm,sglang}. vLLM stores KV in
fixed-size blocks of 16 tokens and reuses a block only if \emph{everything
before it} matches as well, as does SGLang's radix tree~\cite{sglang}; reuse
therefore stops at the first block that differs.

Prompts assembled from shared pieces make this structure graded: retrieved
chunks in RAG, and history and tool outputs in agents. We use RAG as the
example. A RAG prompt is a system prompt $S$, the retrieved chunks, and the
question $q$. Suppose an earlier request $[S \mid A \mid B \mid C \mid q_1]$
left $S$, $A$, $B$ and $C$ in the cache. A new request
$[S \mid A \mid B \mid D \mid q_2]$ reuses $S$, $A$ and $B$, and
recomputes from $D$ onward; $[S \mid D \mid A \mid B \mid q_3]$ reuses
only $S$, even though $A$ and $B$ are cached; and
$[S \mid A \mid B \mid C \mid q_4]$ computes only its question $q_4$. The work a request needs is its number of
\emph{uncached} tokens, and it depends on what the cache holds at the moment
the request is served.
Other systems raise how much of each prompt is cached, by reusing
non-prefix chunks~\cite{cacheblend}, reordering retrieved
chunks~\cite{contextpilot}, or managing agent KV by when each agent runs
next~\cite{kvflow,cachewise}; \sys{} orders the partly cached requests
they produce.
When KV memory is full, an eviction policy decides which blocks survive:
e.g., least-recently-used or rules that protect blocks needed by waiting
requests~\cite{peek}.

\subsection{Scheduling with a Prefix Cache}
\label{sec:bg:sched}

A request computes only the tokens missing from the KV cache when it is
admitted, so the order of the waiting queue decides not only who waits but
also how much reuse each request gets, since a request held back may find
its cached prefix evicted by those admitted before it. First-come-first-served
(FCFS) serves requests in arrival order, regardless of how much of their
prompt is cached. \emph{KV reuse-first} policies rank requests by the extent
of reuse possible: longest-prefix-match (LPM),
offered by SGLang~\cite{sglang}, serves the request with the longest cached
prefix first; RAGCache's cache-aware reordering~\cite{ragcache} ranks by
the KV reuse ratio (cached over uncached tokens). Ranking by reuse alone
raises cache hits and throughput, but a request that keeps losing to
cache-rich newcomers can wait without bound.
Offline systems group requests that share a prefix for throughput, without per-request latency targets~\cite{batchllm,blendserve}.

Existing schedulers add fairness in one of three ways. \emph{Admission counting}:
k-LPM~\cite{klpm} serves the oldest request once every $k$ picks. Preble~\cite{preble} sorts requests into groups by their cache ratio and
gives each group a number of picks that grows with its ratio. Prism reserves a fixed number of first-come slots per batch for rare prefixes~\cite{prism}. \emph{Retention}: PEEK~\cite{peek}
orders requests by the prefill their cached prefix saves and protects the
blocks that waiting requests will reuse from eviction. \emph{Waiting time}:
the classic remedy adds the time a request has waited to its priority, as
PrefillOnly does on top of a ranking by uncached tokens~\cite{prefillonly}, and TOPAS adds task-level aging for agents~\cite{topas}.

Size-aware schedulers favor short jobs: SRPT~\cite{srpt} and
FastServe~\cite{fastserve} at the expense of long ones, and
Autellix~\cite{autellix} favors agent programs that have received less
service so far. Highest response ratio next~\cite{hrrn} measures a
job's wait relative to its service time, the idea behind our relative
wait.

These remedies measure fairness in \emph{requests} (admission counting)
or in \emph{seconds} (waiting time); none measures
fairness in the \emph{work} a pick costs. Section~\ref{sec:mot} shows why: a pick may cost
tens to thousands of tokens(\S\ref{sec:mot:cost}), the best schedule
depends on load (\S\ref{sec:mot:load}); waiting in seconds ignores request size (\S\ref{sec:mot:size}), and no waiting-time rule alone
bounds how much work can be served ahead of a request
(\S\ref{sec:mot:bound}).

{
\section{Motivation}
\label{sec:mot}

We now show why the existing schedulers of \S\ref{sec:bg:sched} fall
short and derive four requirements for a scheduler, using a production
agentic trace and RAG workloads.

\subsection{Cache-Dependent Prefill Work}
\label{sec:mot:cost}

\begin{figure}[t]
  \centering
  \includegraphics[width=\columnwidth]{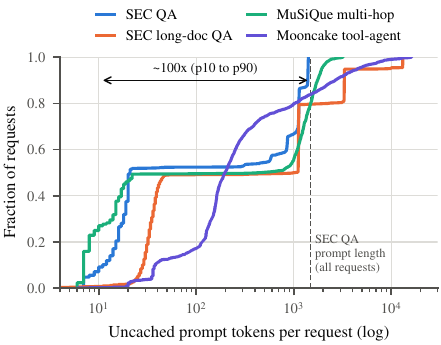}
  \vspace{-5mm} %
  \caption{Uncached prompt tokens per request when each trace is served
  under FCFS (log scale). 
  }
  \label{fig:m1}
  \vspace{-5mm} %
\end{figure}

Graded reuse makes the prefill work of requests in one queue vary widely.
Figure~\ref{fig:m1} shows the tokens the engine actually computes for each
request when each trace is served on GPU: the ratio between the 90th and
the 10th percentile is 70--250$\times$ across our four workloads. Every
SEC QA prompt holds the five chunks retrieved for its question, about
1.5k tokens, yet about half of the requests are full hits that compute
only their question, tens of tokens, while the 30--35\% that are partial
hits carry 54--64\% of the prefill work. The spread comes from the cache,
not from prompt length.

Production traffic shows the same graded reuse. In Mooncake's traces from
Kimi's serving system~\cite{mooncake}, an agent resends its history and
tool outputs on every call, so a request shares a long prefix with earlier
calls of its session and adds a new uncached tail. 58\% of conversation
requests and 65\% of tool-agent requests are partly cached (counting a
block as cached if any earlier request contained it), and their uncached
tokens range from a few hundred to about 18k, a spread of about
$100\times$.

This variation is what makes the queue order matter: serving a full hit
first delays the requests behind it by a few tens of tokens of work,
serving a miss first delays them by thousands, yet a scheduler that counts
admissions or absolute wait time treats the two picks as equal. Prior
evaluations rarely expose this, as they use equal-length
prompts~\cite{peek} at steady load and report mean latency
only~\cite{peek,ragcache}.

{
\subsection{No Fixed Order Fits Every Load}
\label{sec:mot:load}

\begin{figure}[t]
  \centering
  \includegraphics[width=\columnwidth]{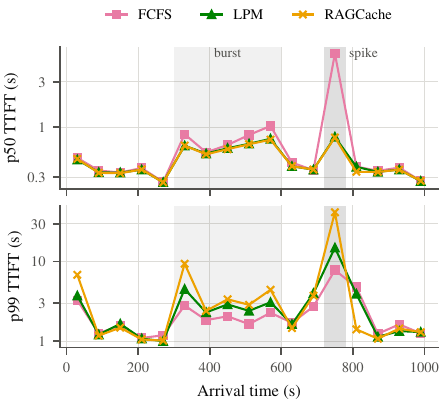}
  \vspace{-9pt} %
  \caption{Median (top) and p99 (bottom) TTFT per 60\,s window for
  the SEC QA topic-shift trace, under FCFS and the two KV reuse-first schedulers (\S\ref{sec:bg:sched}). Shaded: a burst in which 40\% of requests turn to cold
  topics and a 60\,s request spike.}
  \vspace{-9pt}
  \label{fig:m2}
  \vspace{-9pt} %
\end{figure}
\vspace{-3pt}
With prefix caching, a prefill node's load is the prefill work arriving
per second, the request rate times the uncached tokens per request, so it
can rise with more requests or with more KV cache misses.
Figure~\ref{fig:m2} replays a trace that does both: from calm load at 75\%
of the node's capacity, a five-minute burst shifts 40\% of requests to
rarely asked topics, so fewer requests arrive but more of them miss, and
load rises to 93\%; a 60\,s spike in the request rate then pushes load to
114\%. While load is calm, FCFS and the KV reuse-first schedulers are
indistinguishable: queues are short, so the order barely matters. When
load rises, each fails on a different metric. In the spike, FCFS's median
TTFT rises to 5.9\,s, against 0.8\,s for the KV reuse-first schedulers, as
cheap hits wait behind expensive misses. The KV reuse-first schedulers
instead lose the tail: their p99 reaches 15.0\,s and 41.8\,s, against 7.8\,s for FCFS, as cache-rich new requests are
prioritized over the misses.

Schedulers that add fairness to a KV reuse-first order pick a fixed point
between these extremes, k-LPM and Preble by counting admissions and PEEK
by protecting cached blocks. The setting that
keeps the tail short depends on load and on how many requests miss: on these traces,
each stays above FCFS's tail in at least one spike (\S\ref{sec:eval:adapt}). Counting admissions also ignores what each
one costs: one request may cost tens of tokens or thousands
(\S\ref{sec:mot:cost}). Load balancing does not remove such surges: a
demand surge reaches all nodes at once, cache-aware routers deliberately
send requests that share a prefix to the same node~\cite{preble,sglang},
and adding capacity takes far longer than a load spike duration.

\textbf{Requirement R1:} while queues are short, serve the requests with
the most KV cache reuse first; when a burst or spike pushes load toward or
past capacity and requests accumulate in the queue, bound how long a request
can wait while later requests are admitted ahead of it.
}

{
\subsection{Absolute Wait Time Ignores Request Size}
\label{sec:mot:size}

\begin{figure}[t]
  \centering
  \includegraphics[width=\columnwidth]{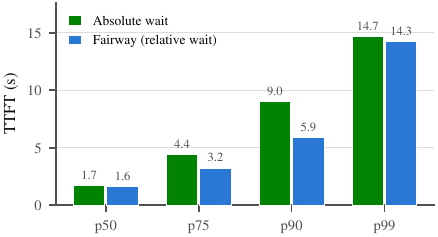}
  \vspace{-5mm} %
  \caption{TTFT percentiles for SEC long-doc QA, under KV
  reuse-first scheduling with an absolute wait and under \sys{}.}
  \vspace{-5mm} %
  \label{fig:m3}
\end{figure}

A common fix for the tail of \S\ref{sec:mot:load} is to add the absolute wait
time to the KV reuse ratio~\cite{ragcache}, and on both dynamic traces it
keeps the KV reuse-first median in calm traffic and FCFS-like tails in the
burst and the spike. Its weakness appears when prompt lengths differ.

The same wait costs requests of different sizes differently. Take two
requests with the same KV reuse ratio, of 1k and 10k tokens, whose
prefill takes about 0.14\,s and 1.4\,s: waiting one second is seven times
the short request's prefill time, but less than the long one's. An
absolute wait ignores this, as every request gains the same priority for
every second it waits. On SEC long-doc QA, whose documents range from
about 1k to 13k tokens, it gives every size about the same p99 TTFT once
a queue forms, 14--16\,s, as do LPM and FCFS, so a short request waits 86
times its own prefill time and a long one 9 times. With near-uniform
prompt lengths, the two kinds of wait behave the same.

A \emph{relative wait}, which divides each request's wait by its own
expected prefill time, favors the request for which waiting is 
more costly, while the reuse ratio still decides between requests whose cache hits differ. On SEC long-doc QA, it lowers the mean TTFT
by 21\% and the p90 by 35\% (Figure~\ref{fig:m3}), and short requests
(62\% of the trace) gain most. Long requests (9\%) pay with a 30\% higher
p99, yet relative to their size they still wait the least; without a
limit on how much shorter work can be served ahead of them, their p99
rises by 47\%, so that work must be bounded (R3).

\textbf{Requirement R2:} judge each request's wait relative to its own
prefill time, so that a short request becomes urgent sooner than a long
one.
}

{
\subsection{Waiting Time Alone Cannot Bound the Wait}
\label{sec:mot:bound}

A relative wait lets short requests leave the queue sooner, but, as
\S\ref{sec:mot:size} showed, it can push long requests back, and it shares
a blind spot with every waiting-time rule: a cached request starts ahead
by its reuse, and an old miss is served before it only after waiting long
enough to make up that head start, so the higher the other requests' KV
cache hits, the longer the miss waits.

This matters when many highly cached requests arrive while misses are
queued: a document suddenly becomes popular, e.g., a filing that was just
released, or an agent's planner issues many tool or sub-agent calls in
parallel~\cite{llmcompiler} that share one long cached
context~\cite{sglang,kvflow}. Each new request enters far ahead of the
waiting misses, so a miss's wait is set by what the other requests look
like, not by its own size or how long it has waited
(\S\ref{sec:eval:flood}). Such bursts may be rare, but a scheduler should
stay safe when they happen, and waiting time alone cannot provide this
protection without tuning: any rule in seconds needs a weight or
threshold that suits one workload and load but not another; too loose,
and the miss still waits; too tight, and the order falls back to FCFS and
loses the median. The oldest requests therefore need a bound measured in
work.

\textbf{Requirement R3:} for oldest requests, keep the
requests served ahead of them within a fixed fraction of their size, no matter how many other requests wait or how
many of them have high KV cache hits.

\textbf{Requirement R4:} meet R1--R3 without per-workload tuning. The mix
of hits, lengths and load is not known in advance and changes over time,
and no fixed trade-off suits every mix (\S\ref{sec:mot:load}).

\textbf{Summary.} The best schedule depends on the mix of cache hits,
prompt lengths and load, and that mix changes over time.
Section~\ref{sec:design} meets R1--R4 with \emph{adaptive reuse
prioritization}: a Fast lane that ranks requests by their KV reuse ratio
plus a relative wait (R1, R2), and a Fair lane that bounds the work served
ahead of the oldest request (R3), with the same parameters on every workload (R4).
}
\begin{figure}[t]
  \centering
  \includegraphics[width=\columnwidth]{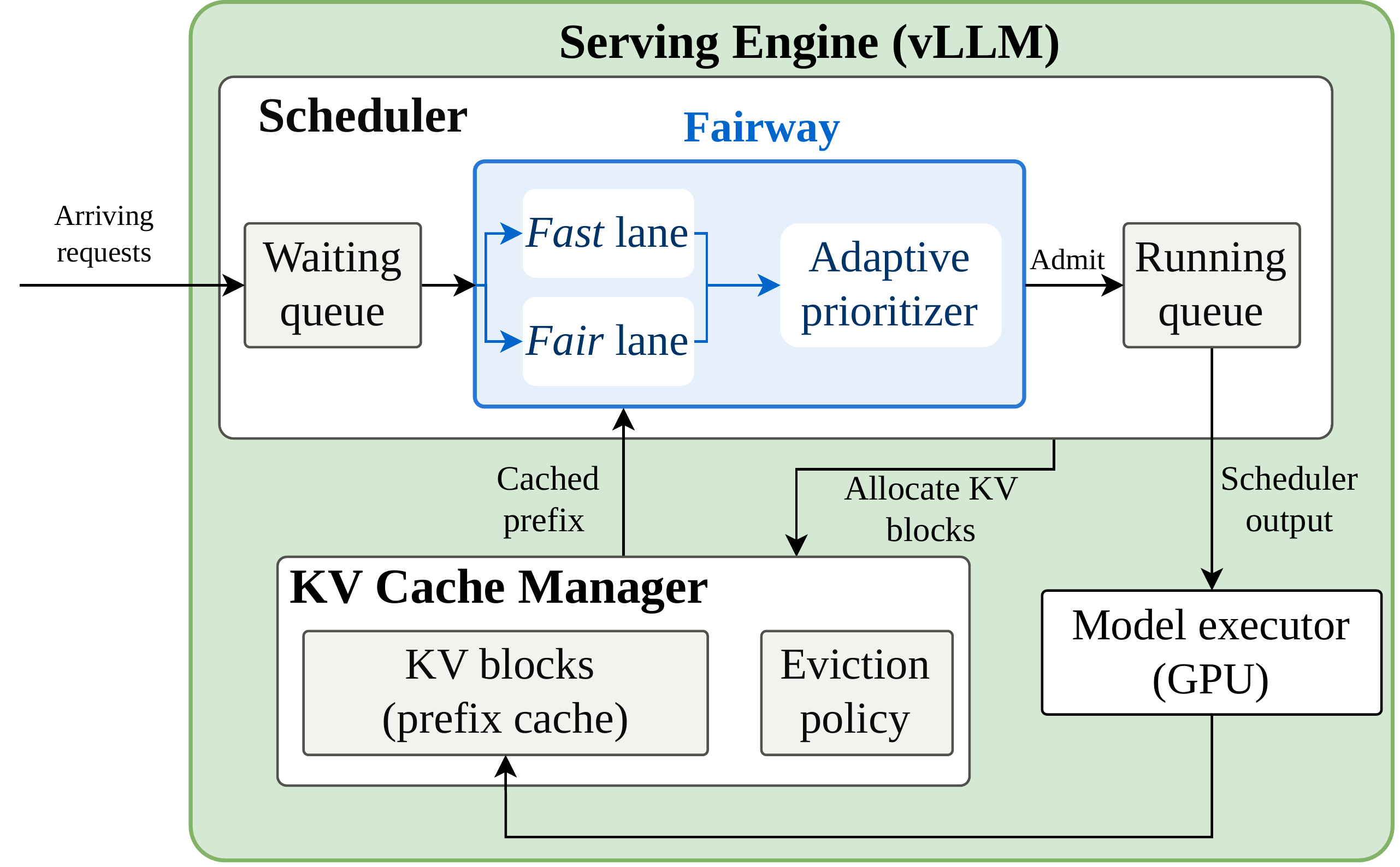}
  \vspace{-8mm} %
  \caption{\sys{} in a serving engine.}
  \vspace{-12pt} %
  \label{fig:arch}
  
\end{figure}

{
\section{\sys{}}
\label{sec:design}

\subsection{System Overview}
\label{sec:design:overview}

\sys{} applies adaptive reuse prioritization in the serving engine's
scheduler (Figure~\ref{fig:arch}) with three components. The
\emph{Fast lane} ranks the waiting requests by KV reuse: it reads each
request's cached prefix from the KV cache manager and scores the request
by the cached tokens it saves per uncached token, plus how long it has
waited relative to its own size. The \emph{Fair lane} ranks the same
requests by arrival time. The \emph{adaptive prioritizer} merges the two
lanes into one ordering based on a running total of the uncached tokens each lane has placed. 
Before each
scheduling step, \sys{} rewrites the order of the waiting queue, so it
reflects the prefixes cached by the previous step, and the step admits
requests in this order. Algorithm~\ref{alg:sched} gives
the full procedure.

These mechanisms meet the requirements of \S\ref{sec:mot}. The Fast
lane's score combines a \emph{reuse-first ranking} (R1,
\S\ref{sec:design:rank}) with a \emph{relative wait} that measures
waiting against each request's own size (R2, \S\ref{sec:design:wait}).
The prioritizer tracks each lane's weighted work as a \emph{work charge}
(\S\ref{sec:design:lane}), which bounds how much work cache hits can place ahead of the oldest request (R3, \S\ref{sec:design:bound}), while the final ordering adapts to each workload's mix of hits, lengths and load (R4,
\S\ref{sec:eval:setup}).

\begin{figure*}[t]
  \centering
  \includegraphics[width=\textwidth]{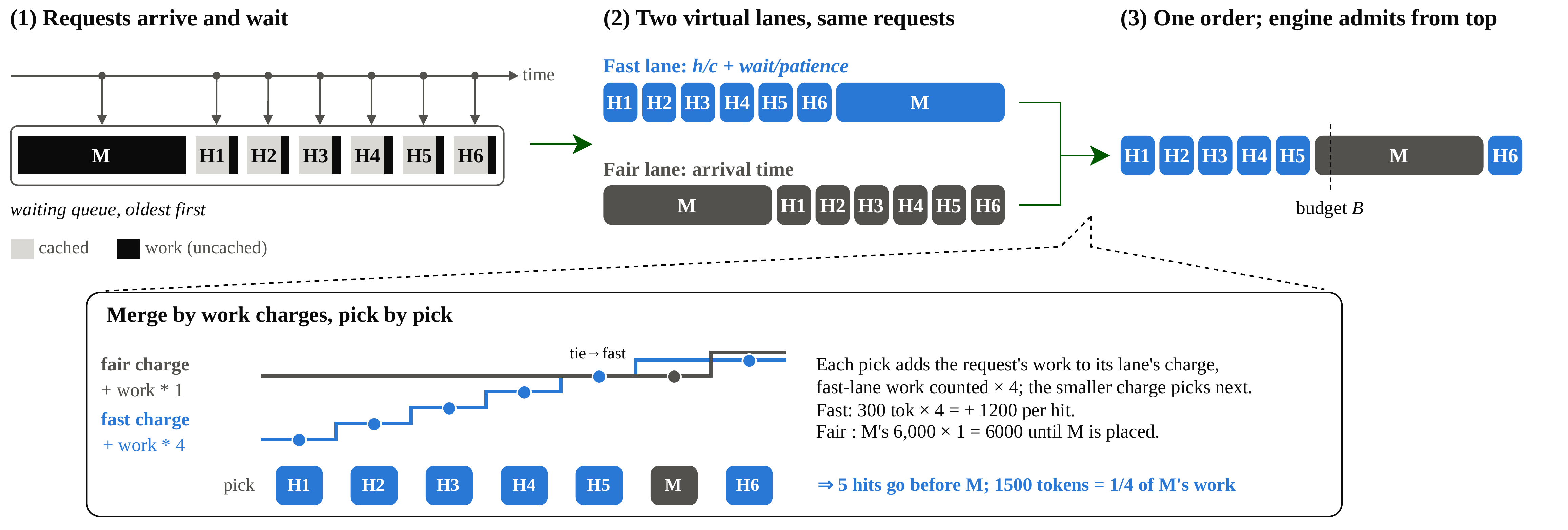}
  \vspace{-5mm} %
  \caption{One \sys{} step with $w=4$: the oldest request $M$ (miss) must compute 6{,}000 tokens; six later hits $H_1$--$H_6$ 300 tokens each. (1)~Waiting queue (2)~Two lanes and their merge by work charges (3)~Step admits from the top of the order within budget $B$.}
  \label{fig:design}
  \vspace{-3mm} %
\end{figure*}

\begin{algorithm}[t]
\caption{\sys{} ordering of the waiting queue, run before every scheduling step.}
\label{alg:sched}
\small
\begin{algorithmic}[1]
\REQUIRE waiting set $Q$, Fast-lane weight $w$ ($=4$), prefill-time estimate $\mathit{est}$
\ENSURE admission order of $Q$
\FORALL{$r \in Q$}
  \STATE $h_r \gets$ cached prefix length of $r$
  \STATE $c_r \gets \max(1,\ \mathit{len}_r - h_r)$
  \STATE $\mathit{SLO}_r \gets \max(1\,\mathrm{s},\ 3\cdot\,\mathit{est}(\mathit{len}_r))$
  \STATE $s_r \gets (h_r / c_r + \mathit{wait}_r / \mathit{SLO}_r)$
\ENDFOR
\STATE $\mathit{Fast} \gets Q$ by $s_r$, highest first (ties: oldest first)
\STATE $\mathit{Fair} \gets Q$ by arrival, oldest first
\STATE $A \gets w\,c_{\mathit{Fast}[1]}$, $F \gets c_{\mathit{Fair}[1]}$ \COMMENT{work charges of the Fast and Fair lanes}
\STATE $\mathit{order} \gets [\,]$
\WHILE{$|\mathit{order}| < |Q|$}
  \IF{$A \le F$}
    \STATE $r \gets$ next unplaced in $\mathit{Fast}$
    \STATE $A \gets A + w\,c_r$
  \ELSE
    \STATE $r \gets$ next unplaced in $\mathit{Fair}$
    \STATE $F \gets F + c_r$
  \ENDIF
  \STATE append $r$ to $\mathit{order}$
\ENDWHILE
\STATE \textbf{return} $\mathit{order}$
\end{algorithmic}
\end{algorithm}

\subsection{Reuse-First Ranking}

\label{sec:design:rank}

For each waiting request $r$ with a prompt of $\mathit{len}_r$ tokens, let
$h_r$ be the length of its cached prefix and
$c_r = \mathit{len}_r - h_r$ the number of uncached tokens it
must compute, its \textit{work}. The Fast lane ranks requests by their
\emph{KV reuse ratio} $h_r / c_r$, the cached tokens a request reuses per
token of work, which RAGCache also uses as its priority~\cite{ragcache}.
Of two requests with the same cached prefix, the ratio puts the one with
less work first, whereas ranking by cached length alone, as done in
SGLang~\cite{sglang}, can put a long partial hit ahead of a cheap full hit
and delay everyone behind it by its larger work (\S\ref{sec:mot:cost}).

\subsection{Relative Wait}
\label{sec:design:wait}
The KV reuse ratio alone starves misses (\S\ref{sec:mot:load}), so the
score adds the time a request has waited, measured relative to its own
expected prefill time: serving a short request first delays the others
little, while serving a long one first delays them a lot
(\S\ref{sec:mot:size}):
\[
  s_r \;=\; \frac{h_r}{c_r} \;+\; \frac{\mathit{wait}_r}{\mathit{SLO}_r},
  \qquad
  \mathit{SLO}_r = \max\bigl(1\,\mathrm{s},\; 3\cdot \mathit{est}(\mathit{len}_r)\bigr),
\]
where $\mathit{est}(\cdot)$ is the expected prefill time of a prompt on an
idle engine, measured once per GPU with a short profiling run. We call the
second term the request's \emph{relative wait} and $\mathit{SLO}_r$ its
\emph{patience}: a request becomes urgent after waiting about three times
its own prefill time, and never sooner than one second. Patience uses the
full prompt length $\mathit{len}_r$, which is fixed for a request, rather
than the work $c_r$, which changes with the cache and already enters the
ratio. Both terms are unitless and are added without a weight, since a
weight on the wait term would only rescale the patience.

When queues are short, waits are a fraction of a second, the relative wait
is small, and the order is KV reuse-first (R1). When a queue forms, the
relative wait grows, fastest for small requests (R2): for the two requests
of \S\ref{sec:mot:size}, the patience is 1\,s and about 4\,s. The
multiplier sets how long KV reuse may delay a request before fairness
takes over. We set it to 3 based on the GPU profiling of
$\mathit{est}(\cdot)$ above, so that a miss that has waited three times its
own prefill time ranks with a newly arrived request that has half its
prompt cached; a smaller multiplier pushes the order toward FCFS and a
larger one toward KV reuse-first.

\subsection{Fair Lane and Work Charges}
\label{sec:design:lane}

A wait term alone cannot bound waiting when many highly cached requests
arrive (\S\ref{sec:mot:bound}). The Fair lane orders the waiting requests by arrival time, oldest first, and the adaptive prioritizer merges it with the Fast lane by work
charges within each step (Figure~\ref{fig:design}). Each lane's charge
starts at the work of its first request, counted $w$ times in the Fast
lane; the Fair lane's first request is the
oldest request $o$, so its charge starts at $c_o$. The lane with the
smaller charge then places its next request and adds that request's work
to its charge, again counted $w$ times in the Fast lane, so each lane's
charge grows only with the work it places. The lanes take turns this way until the whole
queue is ordered, each skipping requests the other has already placed. In
Figure~\ref{fig:design} (inset), five hits go before $M$, 1{,}500 tokens
or a quarter of its work, however many hits wait. The oldest request's own
work thus sets how much cheap work goes ahead of it: about $c_o/w$
(\S\ref{sec:design:bound}).

This merge rule was introduced by stride scheduling ~\mbox{\cite{stride}}: serve
the client with the smallest counter, then advance its counter by the
service it received. Counted up to each of its picks, the Fair lane
receives a share $w/(w+1)$ of the work in the order. The order covers the
whole queue, so this share never leaves a step's budget unused. 

\textbf{Charges restart at every step.} The rule decides how much cheap
work goes before the oldest request in this step, not how work is shared
over time. With $B=2{,}048$ in Figure~\ref{fig:design}, the step admits
the five hits (1{,}500 tokens) and the first 548 tokens of $M$, which is
then running and continues first in the following steps. Carrying the
charges across steps would instead enforce the Fair lane's share (80\% at
$w=4$) of all work over time; under load the queue never empties, so the
order would become close to FCFS and lose the median latency speedup. Restarting the
charges only when the queue empties fails the same way.

\textbf{The bound and its limit.}\label{sec:design:bound} The work placed
ahead of $o$ is at most $c_o/w$ plus one request
(Appendix~\ref{app:bound}); if the budget left in the step covers it and
KV memory is available, $o$ starts its prefill in that step. The bound
acts within one step, so it covers oldest requests of up to about $wB$
uncached tokens, or 8k at $B=2{,}048$. A longer request is still served:
its relative wait keeps growing while it waits, whereas a newly arrived
request scores little more than its ratio, so it eventually rises to the
top of the Fast lane. In our evaluation no request starves, even with
prompts of up to 16k tokens, and \sys{}'s worst wait stays within
2.3$\times$ its p99 (\S\ref{sec:eval}).

}

\begin{figure*}[t]
  \centering
  \includegraphics[width=\textwidth]{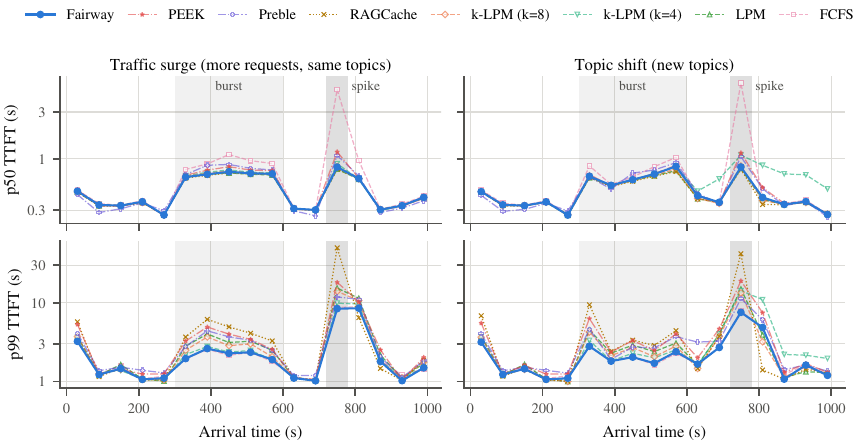}
  \vspace{-5mm} %
  \caption{Median and p99 TTFT per 60\,s window on the two held-out dynamic
  SEC traces (log scale).}
  \label{fig:eval:adapt}
  \vspace{-5mm} %
\end{figure*}

{
\section{Evaluation}
\label{sec:eval}

We evaluate \sys{} under changing load (\S\ref{sec:eval:adapt}), across different workload characteristics (\S\ref{sec:eval:robust}), and under a flood of cached requests (\S\ref{sec:eval:flood}). An ablation attributes each result to one of
\sys{}'s mechanisms (\S\ref{sec:eval:ablation}), and we close with
\sys{}'s overhead and limits (\S\ref{sec:eval:costs}).

\subsection{Setup}
\label{sec:eval:setup}

\paragraph{Platform.} We serve Llama-3.1-8B~\cite{llama3} with vLLM~0.16.0 on
one NVIDIA A6000 (48\,GB) as the prefill node of a disaggregated
deployment~\cite{distserve,splitwise}: each request generates one token, and
we report its TTFT. The token budget is $B=2{,}048$
per step. The KV cache holds a quarter to a third of each trace's
distinct reusable blocks, so eviction is frequent. FCFS is vLLM's default
order. All methods reorder the waiting queue inside vLLM's
scheduler (\S\ref{sec:design:overview}) using the same eviction policy
(Appendix~\ref{app:evict}). Every request reaches the engine at its arrival time and the measured TTFT is the engine's own waiting and prefill time. We report median and p99 TTFT, and call a scheduler \emph{starving} when
its worst TTFT exceeds three times its own p99, so that a low p99 cannot
hide requests that wait minutes; in our runs the border is clear, with
every run either below 2.4$\times$ or above 3.3$\times$.

\paragraph{Workloads.} The RAG workloads are built on the following benchmarks:
SEC 10-Q filings from the KG-RAG datasets~\cite{kgrag}, multi-hop questions
over Wikipedia from MuSiQue~\cite{musique}, and long-document QA from
LongBench and LooGLE~\cite{longbench,loogle}. Each prompt holds the chunks retrieved for
its question (\emph{SEC QA}, MuSiQue) or the whole document (\emph{SEC
long-doc QA}, public QA). What we model is the request process, which
questions arrive and when, as in prior works
~\cite{ragcache,preble,peek}: question popularity follows a Zipf
distribution with $\alpha=0.9$, as in web and RAG
traffic~\cite{breslau,ragpulse}, and arrivals are
Poisson~\cite{vllm,distserve,preble}. Poisson arrivals have no bursts
of their own, so we add them in two ways: the dynamic traces below, and
Mooncake~\cite{mooncake}, a production agent trace with its own bursty
arrivals and prefix sharing.
Table~\ref{tab:eval:workloads} in Appendix~\ref{app:workloads} gives each
workload's details.
\paragraph{Load and dynamic traces.} We set the load of each trace
relative to what the engine can serve: $\rho=\lambda/\mu$, where
$\lambda$ is the arrival rate and $\mu$ the request rate FCFS sustains
when offered more than it can serve, measured before any
comparison. We run each static trace at
$\rho=0.9$ (busy) and $\rho=1.1$ (overloaded), with 4,000 requests (2,932
for Mooncake). Production load is not steady (\S\ref{sec:mot:load}), so
for SEC and MuSiQue we also build two dynamic traces whose load changes
within one run: 300\,s of calm load ($\rho\approx0.75$), a 300\,s burst
($\rho\approx0.9$--1.0), 120\,s of calm, a 60\,s spike past capacity
($\rho\approx1.05$--1.15), and 240\,s of calm, with $\rho$ counted in
uncached tokens (Appendix~\ref{app:workloads}). The two traces differ
in what the burst brings, at about the same work. In the \emph{traffic
surge}, it brings more requests of the same kind. In the \emph{topic
shift}, 40\% of its requests ask questions from the less popular half of
the pool, whose documents are rarely cached; on SEC, this raises the share
of misses from 17\% to 24\%.

\begin{figure*}[t]
  \centering
  \includegraphics[width=\textwidth]{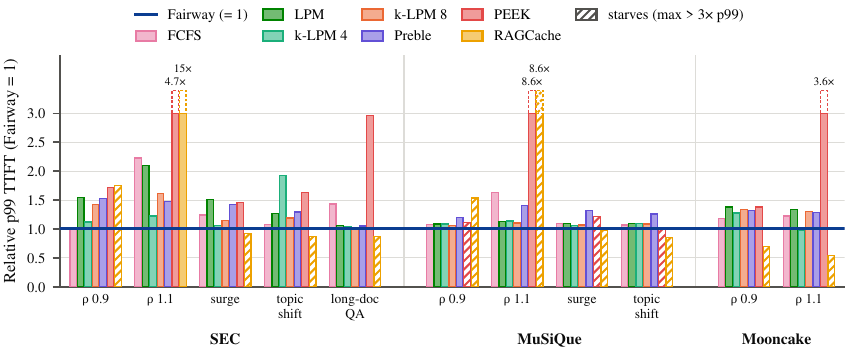}
  \vspace{-7mm} %
  \caption{p99 TTFT of each baseline relative to \sys{}'s (line at 1) in
  all 11 scenarios; above 1 is worse than \sys{}. Hatched bars starve requests
  (worst TTFT $>3\times$ p99).
  Bars above $3\times$ continue dashed, with their value on top.}
  \label{fig:eval:robust}
  \vspace{-3mm} %
\end{figure*}

\paragraph{Baselines.} FCFS; longest prefix match (LPM)~\cite{sglang};
k-LPM~\cite{klpm} with $k=4$ and 8; RAGCache's cache-aware
reordering~\cite{ragcache}; PEEK~\cite{peek}; and Preble's per-GPU
scheduler~\cite{preble}. As an ablation study (\S\ref{sec:eval:ablation}), we
also run partial versions of \sys{} that add its mechanisms one at a time
to demonstrate their effectiveness.

\paragraph{Method.} \sys{} uses three parameters on every workload: the
Fast-lane weight $w=4$, which keeps the work ahead of the oldest request
within about a quarter of its own while still letting several cheap hits
go first (\S\ref{sec:design:bound}); the patience multiplier 3
(\S\ref{sec:design:wait}); and the 1\,s patience floor, without which a
short request's patience would be a fraction of a second and the
ordinary sub-second waits of a busy but stable engine would outweigh the
KV reuse ratio. We
developed and tuned \sys{} on SEC traces drawn with one seed and then
froze it. We then ran it unchanged on SEC traces drawn with a new seed, on
MuSiQue, and on Mooncake. The 11 scenarios are SEC at $\rho=0.9$ and 1.1, both dynamic
traces and long-doc QA; MuSiQue at $\rho=0.9$ and 1.1 and both dynamic
traces; and Mooncake at $\rho=0.9$ and 1.1. Public QA and the flood are
reported separately.

\subsection{Adapting to Changing Load}
\label{sec:eval:adapt}

Figure~\ref{fig:eval:adapt} follows median and p99 TTFT through the two
dynamic traces. No fixed order is best on both: among the orders that do
not starve requests, LPM has the lowest median, and FCFS, which lets no
request overtake another, has the shortest tail at busy load. We therefore
set two targets for each phase (calm, burst and spike): LPM's median and
FCFS's p99, each within 10\%, so six targets per trace. \sys{} meets all
six on both traces, in every run. No baseline meets more than four.

In the spike, the baselines miss in two ways. FCFS keeps its
short tail, but its median rises six- to sevenfold, to 5.1--5.9\,s. The
KV reuse-first orderings keep a low median but lose the tail: their p99
reaches 11--12\,s for Preble, 13--15\,s for LPM and k-LPM 8, 18--19\,s
for PEEK and 42--50\,s for RAGCache. k-LPM 4 stays close to FCFS's tail
in the traffic surge (9.9\,s), but after the topic shift turns many
requests into misses, its p99 rises to 14.2\,s. \sys{} keeps a median of
0.82\,s on both traces and a p99 of 8.4\,s and 7.6\,s. Over the whole run,
where calm time dominates, the medians look alike; the differences sit in
the burst and spike windows.

\sys{} adapts without a mode switch. In calm windows, waits are short, the
relative wait stays small and the order is KV reuse-first, as in LPM. When
a burst builds a queue, the relative wait grows and the oldest requests
move up, keeping the tail near FCFS's, with the same setting throughout.

\subsection{Robustness Across Workloads}
\label{sec:eval:robust}

Figure~\ref{fig:eval:robust} compares p99
TTFT in all 11 scenarios. The best baseline changes from case to case: FCFS
in four scenarios, k-LPM 4 in four and k-LPM 8 in three. \sys{} has a lower
p99 than every baseline that does not starve in 10 cases, by 2--18\%
against the best of them, and ties in the eleventh. SEC shows where the gain comes from. At busy load, \sys{} keeps LPM's
median (0.70\,s against 0.68\,s) and FCFS's tail (a p99 of 3.8\,s against
3.9\,s, while LPM reaches 5.9\,s). Under overload, its p99 of 10.6\,s is
18\% below the best baseline and half of LPM's 22.2\,s.

A p99 can also hide requests that wait far longer than the rest
(hatched in Figure~\ref{fig:eval:robust}). RAGCache's ordering starves requests in 10
scenarios, with worst waits of 16--365\,s, and in the eleventh its p99
reaches 162\,s. PEEK starves requests in three scenarios, and in four more
its p99 exceeds twice FCFS's. Preble never starves, but its p99 is
5--53\% above \sys{}'s in every scenario.
\sys{} starves no request: in every scenario, its worst wait stays within
2.3$\times$ its p99. Among the baselines, k-LPM is the strongest and also
starves no request, but no single $k$ fits every scenario (Appendix~\ref{app:klpm}). A small
$k$ lets the oldest request through often, which works on Mooncake, but
after a topic shift fills the queue with misses, its p99 is 92\% above
\sys{}'s. A large $k$ keeps the cache hot but lets graded reuse push older
requests back (+42--61\% on SEC, +30\% on Mooncake). \sys{} counts work
rather than picks, so what can pass the oldest request scales with that
request's size, with no per-workload setting.

On public QA benchmarks, the
non-starving reuse-aware orders end within 8\% of each other on p99. \sys{} is within 3.3\% of the best (k-LPM 8) and has the
lowest median (tied with LPM) and mean, while RAGCache's ordering starves and the p99
of FCFS and PEEK reaches 116\,s and 1{,}654\,s. Under steady load with long
prompts, the tail is set by how long the queue takes to drain, which no
order can shorten.

\subsection{Bounding the Wait Under a Flood}
\label{sec:eval:flood}

This stress test checks robustness in the worst case for KV reuse-first
ordering (\S\ref{sec:mot:bound}): many highly cached requests arriving
while misses wait. A backlog of misses, built up over a minute, sits in the
waiting queue when a stream of requests with high KV cache hits arrives, each needing only a small prefill. Under LPM, PEEK and RAGCache, each new cheap request passes the
backlog, and the worst miss waits 66--201\,s
(Table~\ref{tab:eval:flood}). \sys{} keeps it at 22\,s, as do FCFS and
k-LPM 4: the Fair lane limits how much cheap work goes ahead of the oldest
miss in each step.

This protection costs the flood requests latency: they wait about as long as
under FCFS (median 18\,s, against 1.6\,s under LPM). In each step, the
Fast lane may place only about a quarter of the oldest miss's work ahead
of it (\S\ref{sec:design:bound}), far less than the flood brings, so
while the backlog lasts \sys{} serves the misses largely in arrival
order, with a few hits between them. The hits are delayed, not starved:
once the backlog clears, they go first again.

\begin{table}[t]
  \centering
  \small
  \caption{Flood stress test, TTFT (s). Misses: backlog requests whose
  prefix is not in the KV cache. KV-cache hits: flood requests whose
  document prefix is already in the KV cache.}
  \label{tab:eval:flood}
  \begin{tabular}{@{}lrrrr@{}}
    \hline
    & \multicolumn{2}{c}{misses (backlog)} & \multicolumn{2}{c}{KV-cache hits (flood)} \\
    scheduler & p50 & max & p50 & p99 \\
    \hline
    \sys{} & 11.7 & 22.0 & 18.2 & 22.2 \\
    FCFS & 11.8 & 22.5 & 18.6 & 22.7 \\
    k-LPM 4 & 11.8 & 22.4 & 18.3 & 22.1 \\
    Preble & 7.8 & 17.2 & 21.9 & 28.2 \\
    LPM & 12.3 & 123.9 & 1.6 & 4.4 \\
    PEEK & 4.7 & 66.3 & 3.1 & 63.5 \\
    RAGCache & 1.4 & 201.4 & 1.2 & 2.0 \\
    \hline
  \end{tabular}
  \vspace{-5mm} %
\end{table}

\subsection{Ablation of \sys{}'s Mechanisms}
\label{sec:eval:ablation}

Table~\ref{tab:eval:ablation} adds \sys{}'s mechanisms one at a time to
the KV reuse ratio, under three stresses: a topic shift, mixed prompt
lengths and the flood. The KV reuse ratio alone has the lowest p99 in the
topic shift, but starves requests to get it (worst wait 62.9\,s). Adding
an absolute wait removes the starvation, and when requests are alike it
is enough: with near-equal prompt lengths it matches \sys{} within 3\%,
and it meets all six targets on both dynamic traces.

The difference appears under stress, when the waiting requests differ in
size. An absolute wait gives every request the same credit per second, so
once a queue persists, a 1k-token request waits behind a 13k-token one;
the relative wait moves small requests up first, which delay others the least. On SEC long-doc QA, this cuts the p99 of short prompts from 13.6
to 8.6\,s; across the scenarios that mix prompt lengths, short
prompts' p99 drops by 23--42\% and long prompts' rises by 6--31\%
(Appendix~\ref{app:length}). Under overload on Mooncake, where requests
differ in both length and cached prefix, it halves the median (3.2 vs.\
6.6\,s) and lowers the mean by a third (5.4 vs.\ 8.3\,s). The largest
requests pay for this: there, \sys{}'s p99 is 27.6\,s against 24.3\,s.
\begin{table}[t]
  \centering
  \small
  \caption{p99 TTFT (s) as \sys{}'s mechanisms are added one at a time
  (the relative wait replaces the absolute one), under three stresses: a topic shift (worst wait in parentheses), mixed
  prompt lengths (SEC; shortest and longest third of prompts), and the
  flood (worst wait of a backlog miss).}
  \label{tab:eval:ablation}
  \setlength{\tabcolsep}{4pt}
  \begin{tabular}{@{}lrrrr@{}}
    \hline
    & topic & \multicolumn{2}{c}{mixed lengths} & flood \\
    order & shift & short & long & worst \\
    \hline
    KV reuse ratio & 5.9 (62.9) & 3.1  & 25.3 & 201.4 \\
    + absolute wait       & 6.7 (7.9)  & 13.6 & 15.1 & 33.4 \\
    + relative wait       & 6.7 (8.1)  & 5.9  & 20.2 & 59.1 \\
    + Fair lane (\sys{})  & 6.7 (7.9)  & 8.6  & 18.1 & 22.0 \\
    \hline
  \end{tabular}
  \vspace{-5mm} %
\end{table}

Removing the Fair lane shows what it adds. The relative wait alone lets
short prompts out faster (5.9\,s), but long prompts wait longer
(20.2\,s), and in the flood the worst miss waits 59\,s: a miss passes a
hit only when its relative wait exceeds the hit's ratio. With the lane, the worst wait drops to
22\,s, below the absolute wait's 33\,s, and long prompts get part of their
tail back (18.1\,s). Each mechanism is needed for one of the stresses we
test.

\subsection{Overhead and Limits}
\label{sec:eval:costs}

\paragraph{Costs.} \sys{} protects the tail first, as SLOs target it, and
in one regime this costs median latency: persistent overload in which
cached requests are only slightly cheaper than misses. On MuSiQue at
$\rho=1.1$, its median is 3.0\,s against LPM's 0.9\,s, for the lowest p99
(12.6 vs.\ 14.2\,s): as the queue persists, the relative wait of older
requests outgrows the small advantage of these partial hits. Without the
wait term, the median falls to LPM's level but the p99 rises by 37\% on
MuSiQue and 12\% on Mooncake.
Where reuse is deep or queues are short-lived, as at busy load and in
bursts and spikes, \sys{} keeps LPM's median.
\paragraph{Overhead and scope.} Ordering the queue takes 0.02\,ms per step
at the median queue (7 requests) and 0.6\,ms at the largest (509), against
steps of 106--355\,ms; with the cached-prefix lookup that every
cache-aware scheduler performs, it stays under 0.1\% of a step at typical
queues and 1--3\% at the concurrency limit. Load and cache size are set relative to the GPU we evaluate on, so the comparison
carries over to larger GPUs serving larger workloads. At $B=2{,}048$, the
Fair lane admits an oldest request of up to about 8k uncached tokens
within a step, and longer requests are served
through the relative wait; a larger token budget extends this range.
}

{
\section{Conclusion and Future Work}
\label{sec:conclusion}

Prefix caching makes prefill work graded: the uncached work of requests
in one queue can differ by up to $100\times$, and the order in which they
are served sets their TTFT. KV reuse-first orders, FCFS and request-count
fairness each fit some mix of KV cache hits, prompt lengths and load, and
fail on another. \sys{} is built on adaptive reuse prioritization. Its
Fast lane ranks requests by KV reuse and promotes them as they wait longer
relative to their own expected prefill time; its Fair lane, merged with
the Fast lane by work charges, caps the work that cached requests can place ahead
of the oldest request. Without
per-workload tuning, \sys{} matches or beats the best baseline in every
scenario we evaluate, with a p99 TTFT up
to $8.6\times$ lower than that of prior prefix-aware schedulers. In load
spikes, it keeps the tail of FCFS at a $6\times$ lower median TTFT. Future
work includes extending the bound to requests too large to start within
one step, evaluating larger models on multi-GPU prefill instances, and
ordering requests jointly with the cluster router and the eviction policy.

}

\bibliography{example_paper}
\bibliographystyle{mlsys2025}

{
\section*{AI Usage Disclosure}
We used AI coding and writing assistants to speed up parts of this work:
writing scripts to run experiments, collecting and visualizing the evaluation results, porting and running the baselines in
our evaluation framework, and proofreading the manuscript. All
design decisions, experiment plans and results were defined and checked by
the authors, who reviewed and revised the final version of the manuscript and
take full responsibility for its content.
}
\appendix
{
\section{Bound on the Work Ahead of the Oldest Request}
\label{app:bound}

Consider the order computed for one step (Algorithm~\ref{alg:sched}). Let
$o$ be the oldest waiting request and $c_r \ge 1$ the work (uncached
tokens) of request $r$. The Fast lane's charge $A$ starts at
$w\,c_{a_1}$, where $a_1$ is the Fast lane's first request, and grows by
$w\,c_r$ with each Fast pick $r$. The Fair lane's first request is $o$, so
its charge $F$ starts at $c_o$. The Fast lane picks while $A \le F$.

\begin{lemma}
\label{lem:bound}
Let $a_1,\dots,a_k$ be the requests placed before $o$, and
$W = \sum_i c_{a_i}$ their total work. Then
$W \le c_o/w + c_{a_k} - c_{a_1}$: at most $c_o/w$ plus one request,
however many requests wait and however cached they are.
\end{lemma}

\begin{proof}
The Fair lane's first pick is $o$, so every request before $o$ is a Fast
pick, and $F = c_o$ until $o$ is placed. The Fast lane picks $a_k$ only if
$A \le F$ just before, when $A = w\,(c_{a_1} + W - c_{a_k})$. Thus
$w\,(c_{a_1} + W - c_{a_k}) \le c_o$, which gives the bound. The proof uses
only the charges, so the bound holds whatever the scores are.
\end{proof}

When the requests before $o$ have equal work, $c_{a_k} = c_{a_1}$ and the
bound is $c_o/w$: a quarter of $o$'s work at $w=4$
(Figure~\ref{fig:design}, inset).

\begin{corollary}
\label{cor:admit}
$o$ starts its prefill in a step if (i) the token budget left after the
running requests exceeds $c_o/w + c_{a_k}$, (ii) the KV cache can hold the
requests before $o$ and $o$'s first chunk without evicting their cached
prefixes, and (iii) the limit on requests per step is not reached.
\end{corollary}

\begin{proof}
The scheduler admits requests in this order while budget remains. Under
(i)--(iii), all requests before $o$ fit (Lemma~\ref{lem:bound}), and at
least $c_{a_1} \ge 1$ token of budget remains for $o$'s first chunk.
\end{proof}

\paragraph{Scope.} The bound holds within one step; charges restart every
step, so it is not a share of the GPU over time. It lets $o$ start within
the step for oldest requests of up to about $wB$ uncached tokens, 8k at
$B=2{,}048$. A larger request relies on the relative wait, which keeps
growing while a newly arrived request scores little more than its ratio,
so it still reaches the top of the Fast lane. In our evaluation no request
starves, and \sys{}'s worst wait stays within 2.3$\times$ its p99
(\S\ref{sec:eval}).

\section{Workloads}
\label{app:workloads}

\begin{table}[t]
  \centering
  \small
  \caption{Workloads. RAG traces use Zipf popularity ($\alpha=0.9$) and
  Poisson arrivals; Mooncake keeps its own arrivals and sharing.}
  \label{tab:eval:workloads}
  \setlength{\tabcolsep}{4pt}
  \begin{tabular}{@{}lp{3.6cm}r@{}}
    \hline
    workload & content & tokens \\
    \hline
    SEC QA & 20 SEC 10-Q filings~\cite{kgrag} in 256-token chunks; top 5 retrieved & 1.2--1.5k \\
    SEC long-doc QA & same filings, whole documents of three sizes & 1.0--13.2k \\
    MuSiQue & 21k Wikipedia paragraphs~\cite{musique}; top 11 retrieved & 0.9--3.4k \\
    Mooncake & production agent requests~\cite{mooncake} & 0.3--16k \\
    Public QA & LongBench, LooGLE QA at natural length~\cite{longbench,loogle} & 3.5--16.3k \\
    Flood & SEC text; miss backlog, then cached hits & 1.1--6.1k \\
    \hline
  \end{tabular}
\end{table}

\paragraph{RAG traces.} The documents and questions come from the
benchmarks, and each prompt holds the chunks retrieved for its question, or
the whole document for SEC long-doc QA and public QA. Each request asks a
question from a fixed pool, and the $i$-th most popular question is asked
with probability proportional to $i^{-\alpha}$. Requests for web content
follow such Zipf-like distributions with an exponent below
one~\cite{breslau}, and RAG retrievals concentrate on a small set of hot
chunks~\cite{ragpulse}. A larger $\alpha$ turns more requests into hits,
which favors KV reuse-first schedulers, and a smaller one turns more into
misses, where every order approaches FCFS. Popularity ranks are a seeded
shuffle of the pool, so a question's popularity is independent of its
length.

\paragraph{Mooncake.} The trace gives arrival times, sessions, and which
requests share which prefix blocks; we fill the blocks with fixed text, so
that requests share the blocks the trace lists. To fit one GPU, we keep one
in eight sessions, each with all its requests, and scale prompt lengths by
1/4. Its arrivals are bursty (an index of dispersion of 4.5 over 10\,s
windows, against 1 for Poisson). Its native rate is $\rho=0.8$; we compress
time uniformly to $\rho=0.9$ and 1.1.

\paragraph{Load.} The static runs at $\rho=0.9$ and 1.1 model a busy
engine and a sustained but temporary overload. Below about $\rho=0.8$ queues rarely
form and every order gives the same TTFT; above about 1.2 the backlog
grows so fast that every order fails; in between, the order decides who
waits. In the dynamic traces, $\rho$ counts uncached tokens rather than
requests: a burst can change how much uncached work each request brings,
so each phase's request rate is set so that its uncached tokens per second
reach the target fraction of what FCFS serves at capacity. The calm
between phases lets the queue drain, so each phase is measured on its own.
}

\section{Eviction Policy}
\label{app:evict}
When KV memory is full, all methods use PEEK's demand-aware
eviction~\cite{peek}. Blocks that no waiting request needs are evicted
first, least recently used first. Among blocks that waiting requests need,
the policy evicts first those whose loss forces the fewest tokens to be
recomputed, summed over the waiting requests that need them; these are the
ends of cached prefixes, since evicting a block also makes every later
block of its prefix unusable. With eviction, the handling of running
requests, and preemption the same everywhere, the methods we compare are
admission policies: they differ only in which waiting requests are
admitted next, so the differences in our results come from this order
alone.

\section{k-LPM Across Scenarios}
\label{app:klpm}

\begin{table}[t]
  \centering
  \small
  \caption{p99 TTFT (s) of \sys{}, and how much higher k-LPM's p99 is
  in the same scenario. No fixed $k$ fits every scenario.}
  \label{tab:eval:klpm}
  \setlength{\tabcolsep}{8pt}
  \begin{tabular}{@{}lrrr@{}}
    \hline
    & \sys{} & \multicolumn{2}{c}{k-LPM} \\
    scenario & p99 & $k=4$ & $k=8$ \\
    \hline
    SEC, $\rho=0.9$     & 3.82  & +12\% & +42\% \\
    SEC, $\rho=1.1$     & 10.55 & +22\% & +61\% \\
    Traffic surge       & 6.44  & +6\%  & +14\% \\
    Topic shift         & 6.71  & +92\% & +19\% \\
    SEC long-doc QA   & 14.26 & +4\%  & +3\%  \\
    MuSiQue, $\rho=1.1$ & 12.62 & +14\% & +10\% \\
    Mooncake, $\rho=1.1$& 27.57 & \textminus2\% & +30\% \\
    \hline
  \end{tabular}
\end{table}

Table~\ref{tab:eval:klpm} compares \sys{} with k-LPM at $k=4$ and 8 in the
scenarios where the choice of $k$ matters. A small $k$ fails after a
topic shift (+92\%), and a large $k$ fails under steady graded reuse
(+42--61\% on SEC, +30\% on Mooncake). Choosing $k$ thus requires
knowing the workload in advance, and the workload changes.

\begin{table}[h]
  \centering
  \small
  \caption{p99 TTFT (s) of short and long prompts under \sys{}, and its change
  against the KV reuse ratio with absolute wait. Short and long are the shortest and
  longest third of each trace's requests by prompt length (split points in
  tokens); Mooncake at $\rho=0.9$ and 1.1.}
  \label{tab:eval:length}
  \setlength{\tabcolsep}{3pt}
  \begin{tabular}{@{}llrrrr@{}}
    \hline
    & & \multicolumn{2}{c}{short prompts} & \multicolumn{2}{c}{long prompts} \\
    scenario & split & p99 & change & p99 & change \\
    \hline
    SEC long-doc QA  & 1.2k / 3.4k  & 8.6  & \textminus37\% & 18.1 & +20\% \\
    Mooncake 0.9 & 1.2k / 1.8k  & 8.2  & \textminus23\% & 13.2 & +11\% \\
    Mooncake 1.1 & 1.2k / 1.8k  & 14.2 & \textminus42\% & 32.5 & +31\% \\
    Public QA        & 7.5k / 12.6k & 21.4 & \textminus23\% & 31.4 & +6\% \\
    \hline
  \end{tabular}
\end{table}
\section{Short and Long Prompts Under Mixed Lengths}
\label{app:length}

Table~\ref{tab:eval:length} compares \sys{} with the KV reuse ratio plus an
absolute wait, which differ only in how waiting is measured. Short prompts'
p99 drops by 23--42\%, and long prompts', which have more patience, rises
by 6--31\%. On SEC long-doc QA, \sys{} also keeps the lowest overall p99
(14.3 vs.\ 14.7\,s), and its short-prompt p99 is 35\% below that of the
best baseline (k-LPM 8, 13.2\,s). LPM, k-LPM and Preble give the same p99
to all three length groups, like the absolute wait. The exception is
Preble on Mooncake, whose group order serves short prompts faster still
(4.5 and 7.2\,s), at an overall p99 28--32\% above \sys{}'s.

\end{document}